# Negotiating Privacy with Smart Voice Assistants: Risk–Benefit and Control–Acceptance Tensions

Molly Campbell  
*Computer Science Department*  
*Vancouver Island University*  
Nanaimo, Canada  
molly.campbell@viu.ca

Mohamad Sheikho Al Jasem  
*Computer Science Department*  
*Vancouver Island University*  
Nanaimo, Canada  
mohamad.sheikhoaljasem@viu.ca

Ajay Kumar Shrestha  
*Computer Science Department*  
*Vancouver Island University*  
Nanaimo, Canada  
ajay.shrestha@viu.ca

*Abstract*— Smart Voice assistants (SVAs) are widely adopted by youth, yet privacy decision-making in these environments is often characterized by competing considerations rather than clear-cut preferences. While our prior research has examined privacy risks, benefits, trust, and self-efficacy as distinct predictors of behavior, less attention has been paid to how these factors combine into higher-level tension that shapes privacy outcomes. This study introduces a negotiation-based framework for understanding youth privacy decision-making with SVAs by operationalizing two composite indices: the Risk-Benefit Tension Index (RBTI) and the Control-Acceptance Tension Index (CATI), using survey data from 469 Canadian youth aged 16-24. We examine the distribution of these indices and their relationship with privacy-protective behavior and SVA usage. Results show that both indices are meaningfully associated with protective action. Frequent SVA usage exhibits more benefit-dominant and acceptance-leaning negotiation profiles, suggesting that convenience-driven engagement may come at the expense of perceived control. By reframing privacy decision-making as a process of negotiation rather than inconsistency, this study offers a complementary perspective on the privacy paradox and provides a compact measurement approach for capturing how youth navigate competing privacy pressures in voice-enabled ecosystems.

## I. Introduction

Smart Voice assistants (SVAs) such as Apple Siri, Amazon Alexa, and Google Assistant have become embedded in the everyday routines of young people. These systems are used for a wide range of tasks, including information seeking, entertainment, communication, and smart-home control. These systems operate in intimate and shared spaces, often listening passively and logging interactions over time [1], [2]. For youth, using SVAs involves an ongoing negotiation: they weigh perceived privacy risks against the convenience and personalization benefits, and they balance a desire for control with acceptance of platform defaults [3].

Our prior research has identified key constructs that shape privacy decision-making, including perceived privacy risk, perceived privacy benefits, trust and transparency in data practices, and privacy self-efficacy [1]. Structural models have demonstrated how these constructs relate to privacy-protective behavior in online and mobile environments, including SVAs [1], [4]. However, these approaches typically treat each construct separately. They do not directly capture the tensions that arise when risks and benefits pull in opposite directions, or when feelings of control conflict with the temptation to accept convenient defaults. This approach, however, can obscure the fact that youth often experience privacy not as a set of independent evaluations, but as a process of negotiation between competing considerations. Perceived risks may coexist with strong perceived benefits, and a desire for control may conflict with ease of accepting a default setting [5]. These tensions are particularly remarkable in SVA use, where opting out is rarely straightforward and where disengagement may come at the cost of losing valued functionality [6], [7], [8].

In this paper, we move beyond construct-level analysis to adopt a negotiation-based view on youth privacy decision-making. Using survey data from Canadian youth smart voice assistant users, we introduce two composite negotiation indices that summarize how key privacy constructs interact as tensions rather than as isolated factors [1]. The Risk–Benefit Tension Index (RBTI) captures the balance between perceived privacy risks and perceived benefits, while the Control–Acceptance Tension Index (CATI) reflects the balance between control-oriented factors such as privacy self-efficacy, trust in data practices, and benefit-driven acceptance. These indices provide compact, interpretable measures of how youth weigh competing considerations when engaging with SVAs. This study is guided by three research questions: (1) RQ1: Can youth privacy attitudes toward SVAs be summarized into negotiation indices that meaningfully combine existing constructs?, (2) RQ2: How are these indices distributed in the youth population, and what tension patterns do they reveal? and (3) RQ3: How do the negotiation indices relate to privacy-protective behavior and usage patterns, such as frequency of SVA use?

The remainder of the paper is organized as follows: Section II provides background and related works. Section III describes the methodology. Section IV presents the results. Section V provides a discussion. Section VI concludes the paper.

## II. Background and Related Work

### A. Privacy Calculus and Risk-Benefit Trade-offs

Privacy calculus theories suggest that individuals make privacy decisions by weighing perceived technology benefits against perceived risks [1], [4], [6], [9]. Benefits may include convenience, efficiency, personalization, and social value, while risks may involve surveillance, unauthorized access, or secondary use of personal information [9]. In SVAs, this trade-off is especially pronounced: they offer convenient and personalized functionality attractive to youth, while also introducing risks related to always-listening microphones, voice data storage, and third-party sharing. Prior work has treated risks and benefits as distinct constructs and shown that both shape continued use and selective protective actions [1], [4], [10]. However, examining them separately may miss how youth experience the tension between them when deciding whether and how to engage with SVAs.

*B. Trust, Self-Efficacy, and Acceptance of Defaults*

Beyond risk and benefit, privacy decision-making also depends on users' trust in service providers and confidence in managing privacy controls [1], [3]. Algorithmic transparency and trust shape whether users see data practices as understandable, fair, and aligned with their interests. Privacy self-efficacy reflects users' perceived ability to navigate settings, understand permissions, and act to protect personal information [1], [11]. In SVA ecosystems, these factors are shaped by interface design and default settings, which are often buried in menus, expressed in technical language, or enabled by default. As a result, youth may accept data practices not because they endorse them, but because they feel uncertain about intervening or trust providers to act responsibly [1]. This reveals a tension between control and acceptance that is not fully captured when trust or self-efficacy is examined alone [3].

*C. Privacy Paradox as Negotiation and Ambivalence*

The privacy paradox literature describes situations in which individuals express privacy concern yet continue using technologies that collect substantial personal data. While early work framed this as inconsistency or irrationality, more recent research interprets it as negotiation, ambivalence, and constraint [4], [6]. Users may value benefits, recognize risks, and still feel limited in their ability to act, producing behavior that appears contradictory when examined through single constructs. For youth using SVAs, these dynamics are intensified by ubiquitous voice features and everyday integration. Rather than treating concern and use as opposites, a negotiation perspective suggests that youth balance competing pressures in internally coherent ways, even if these appear paradoxical from the outside [3], [4].

*D. Research Gap*

Although prior research has modeled perceived risk, benefits, trust, and self-efficacy as predictors of privacy behavior, limited work has combined these constructs into higher-order measures that explicitly represent negotiation or tension [1], [4], [5], [9]. Existing models support integrating positive and negative bases of attitudes, but rarely offer simple indices that capture how opposing forces, such as risk versus benefit or control versus acceptance, coexist within individuals [1], [12], [13], [14]. This paper addresses that gap by introducing two negotiation indices, RBTI and CATI, that combine established privacy constructs into interpretable measures of risk–benefit and control–acceptance tension. By examining their distribution and their relationship to privacy-protective behavior and SVA use, we provide initial evidence of their validity in the youth SVA context and offer a compact complement to construct-level models of privacy decision-making.

III. METHODS

*A. Dataset and Participants*

This study uses the same cross-sectional survey dataset of Canadian youth described in [15]. Participants were 16 to 24 years old, lived in Canada, and had used an SVA (e.g., Siri, Alexa) at least once in the six months before the survey. Recruitment occurred through university channels, social media, flyers, personal networks, and Canadian school districts. The survey and all instruments received ethics approval from the Vancouver Island University Research Ethics Boards (Ref #103597). After data cleaning to remove responses that did not meet demographic criteria or had substantial missing data, the final analytic sample was N = 469. The sample included 37.1% women, 51.4% men, 3.2% non-binary or another gender identity, 7.5% who preferred not to say, and 0.9% blank or missing responses. The mean age was 18.65 years. Most participants were currently in high school or reported high school as their highest level of education (59.3%). SVA use was reported as rare by 40.5%, daily by 26.9%, weekly by 24.1%, and monthly by 8.1%. Participant demographics are summarized in Table I.

*B. Indices Design & Measures*

The survey measured five core SVA privacy constructs adapted from prior literature: (1) Perceived Privacy Risk (PPR), covering concerns about data collection, covert recording, unauthorized access, and voice-data storage [16], [17]; (2) Perceived Privacy Benefits (PPBf), reflecting perceived usefulness, convenience, and personalization benefits [18], [19]; (3) Algorithmic Transparency and Trust (ATT) , concerning the clarity and fairness of SVA data practices and trust in providers [20], [21]; (4) Privacy Self-Efficacy (PSE), , reflecting confidence in understanding and managing privacy settings and permissions [22], [23]; (5) Privacy-Protective Behavior (PPB), capturing self-reported actions to limit data collection and manage voice-related data, such as reviewing permissions, deleting history, and disabling microphones [24], [25]. Each construct was measured with four items on a five-point Likert scale (1 = "Strongly disagree" to 5 = "Strongly agree"), with higher scores indicating higher levels of the construct. All five four-item scales showed acceptable internal consistency in this sample (Cronbach's α = .88 for PPR, .87 for PPBf, .80 for PSE, and .71–.72 for ATT and PPB).

TABLE I. PARTICIPANT DEMOGRAPHICS

| Characteristic | n | % |
|---|---|---|
| *Gender* | | |
| Blank/Missing | 4 | 0.9 |
| Female | 174 | 37.1 |
| Male | 241 | 51.4 |
| Non-binary/Other | 15 | 3.2 |
| Prefer not to say | 35 | 7.5 |
| *Education Level* | | |
| High School | 278 | 59.3 |
| Post-Secondary | 183 | 39 |
| Blank/Missing | 8 | 1.7 |
| *Frequency of SVA use* | | |
| Daily | 126 | 26.9 |
| Monthly | 38 | 8.1 |
| Rarely | 190 | 40.5 |
| Weekly | 113 | 24.1 |
| Blank/Missing | 2 | 0.4 |
| | **Mean (SD)** | |
| *Age* | 18.65 (2.30) | |

To capture higher-order tensions in privacy attitudes, two negotiation indices were constructed from standardized construct scores, following standard composite-indicator practice [26]. We standardized each construct score to a z-score (mean 0, standard deviation 1) within the sample. For example, z(PPR) = (PPR−μPPR)/σPPR, where μPPR and σPPR are the sample mean and standard deviation of the four-item PPR scale, respectively. The same transformation was applied to PPBf, ATT, and PSE. All indices reported below are based on these standardized construct scores.

RBTI operationalizes the negotiation between perceived privacy threats and rewards. It is calculated as the difference between standardized risk and benefit scores:

$$\text{RBTI} = z(\text{PPR}) - z(\text{PPBf}) \quad (1)$$

Higher RBTI values indicate that perceived risks outweigh perceived benefits (risk-dominant), whereas lower or negative values indicate that perceived benefits outweigh risks (benefit-dominant).

The CATI captures the tension between users' sense of control and their acceptance of benefits for convenience. We first define a Control composite as the average of standardized privacy self-efficacy and algorithmic transparency and trust scores:

$$\text{Control} = \frac{z(\text{PSE}) + z(\text{ATT})}{2} \quad (2)$$

The CATI is then defined as:

$$\text{CATI} = \text{Control} - z(\text{PPBf}) \quad (3)$$

Higher CATI values indicate that control-related factors outweigh perceived benefits (control-dominant), whereas lower or negative values indicate that benefit-driven acceptance outweighs perceived control (benefit-leaning acceptance). As an initial construct validity check, we examined how the indices relate to their component constructs. As expected, RBTI was positively correlated with perceived privacy risk and negatively correlated with perceived privacy benefits, whereas CATI was positively correlated with privacy self-efficacy and algorithmic transparency and trust, and negatively correlated with perceived privacy benefits. These patterns support interpreting RBTI as a risk–benefit balance and CATI as a control–acceptance balance. Their predictive validity in relation to protective behavior and SVA usage is further assessed in Section IV.

### C. Analysis Procedure

We addressed RQ1–RQ3 in three steps. First, for RQ1 and RQ2, we examined the distributions of RBTI and CATI by reporting their means, standard deviations, and ranges, and visualized their relationship with a scatterplot showing how the youth sample is distributed across this two-dimensional tension space. Second, for RQ3, we computed Pearson correlations among RBTI, CATI, and PPB, and estimated linear regression models with PPB as the dependent variable and RBTI and CATI as predictors, controlling for age and SVA usage frequency. Usage frequency was coded on a four-point ordinal scale (1 = Rarely, 2 = Monthly, 3 = Weekly, 4 = Daily). This allowed us to test whether more risk-dominant or control-dominant attitudes are associated with greater protective behavior. Third, to further address RQ3, we analyzed how RBTI and CATI vary by behavioral pattern. We compared mean index values across SVA usage-frequency categories (e.g., daily vs weekly) using one-way ANOVA with Tukey's HSD post-hoc tests, and examined Pearson correlations of both indices with self-reported usage and age. All analyses were conducted in R version 4.4.2.

## IV. RESULTS

### A. Distribution of Negotiation Indices

To address RQ1 and RQ2, we first examined the descriptive statistics for the two negotiation indices (Table II). Both indices are standardized z-score differences, resulting in means of approximately zero and standard deviations near one. On average, RBTI was M = 0.00 (SD = 1.63), with values ranging from -4.98 to 3.63. The distribution was approximately symmetric (Skew = 0.05), indicating that many youths occupy intermediate positions where perceived risks and benefits are comparable. The CATI had a mean of M = 0.00 (SD = 1.01), with a range from -2.58 to 3.19. The distribution was slightly positively skewed (Skew = 0.24), reflecting a modest right tail of respondents with strongly control-dominant positions. Overall, however, youth were split between control-leaning and benefit-leaning attitudes, with a small majority falling on the benefit-leaning side (CATI < 0). Fig. 1 further addresses RQ2 by depicting the joint distribution of RBTI and CATI. Each point represents a respondent in a two-dimensional tension space.

### B. Relationship with Privacy-Protective Behavior

Addressing RQ3, Pearson correlations, as shown in Table III, showed that RBTI and CATI were meaningfully associated with PPB. Youth with higher RBTI (more risk-dominant attitudes) tended to report higher levels of PPB ($r = .26$; $p < 0.001$), while higher CATI (more control-dominant attitudes) were associated with higher PPB ($r = .29$; $p < 0.001$). These patterns suggest that both risk-benefit and control-acceptance tensions are relevant for understanding who engages in protective actions. In the regression model with PPB as the dependent variable, RBTI and CATI, along with control variables, jointly explained a significant portion of variance in behavior $R^2 = 0.133$, $F(4, 406) = 15.57$, $p < 0.001$. This model was estimated on the subset of participants with complete data on PPB, age, and usage frequency (n = 411). Controlling for age and SVA usage frequency, CATI remained a significant positive indicator of PPB (β = 0.209; p < 0.001), indicating that youth who feel more in control than influenced by benefits are more likely to take privacy protective actions. RBTI had a slightly smaller positive effect on PPB (β = 0.074; p = 0.006), suggesting that a heightened sense of risk relative to benefit predicts more protective behavior. Age was a small, positive, and significant predictor (β = 0.032; p = 0.048), while SVA usage was not significant (β = 0.058; p = 0.089). Table IV presents the full regression results.

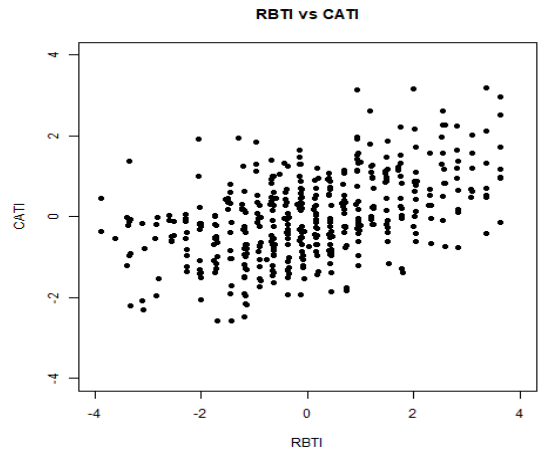

Fig. 1. Scatter plot of Indices

### C. Association with SVA Usage

Further addressing RQ3, we examined how negotiation indices varied by SVA usage patterns. A one-way ANOVA revealed that RBTI values differed significantly by usage frequency, $F(3, 463) = 44.76$, $p < 0.001$. Post-hoc Tukey tests indicated that youth who reported daily use of SVAs had significantly lower RBTI values (M = -0.95; SD = 1.31) than less frequent users (e.g., Weekly: M = -0.41; Rarely: M = 0.88), suggesting that heavy SVA users are more benefit-dominant. A similar pattern was found for CATI, $F(3, 463) = 36.93$, $p < 0.001$, with daily users reporting significantly lower CATI (M = -0.54, SD = 0.83) than rare users (M = 0.48, SD = 1.03), indicating that frequent users experience lesser relative control despite their frequent engagement. Results are summarized in Table V.

TABLE II. DESCRIPTIVE STATISTICS FOR NEGOTIATION INDICES

|  | Mean | SD | Min | Max | Range | Skew |
|---|---|---|---|---|---|---|
| **RBTI** | 0.00 | 1.63 | -4.98 | 3.63 | 8.61 | 0.05 |
| **CATI** | 0.00 | 1.01 | -2.58 | 3.19 | 5.77 | 0.24 |

TABLE III. PEARSON CORRELATIONS

|  | PPB | Age | Usage Frequency |
|---|---|---|---|
| **RBTI** | .26*** | .21*** | -.47*** |
| **CATI** | .29*** | .00 | -.44*** |

TABLE IV. LINEAR REGRESSION RESULTS

| Variable | Coefficient | SE | t | p |
|---|---|---|---|---|
| **RBTI** | 0.074 | 0.027 | 2.75 | 0.0062** |
| **CATI** | 0.209 | 0.042 | 5.01 | < 0.001 *** |
| **Age** | 0.032 | 0.016 | 1.98 | 0.0483* |
| **Usage Frequency** | 0.058 | 0.034 | 1.71 | 0.0886 |
| **Intercept** | 2.316 | 0.303 | 7.66 | < 0.001 *** |

TABLE V. GROUP COMPARISONS OF RBTI AND CATI BY SVA USAGE FREQUENCY

| Usage Frequency | n | RBTI Mean (SD) | CATI Mean (SD) |
|---|---|---|---|
| **Daily** | 126 | -0.95 (1.31) [a] | -0.54 (0.83) [a] |
| **Weekly** | 113 | -0.41 (1.29) [ab] | -0.29 (0.82) [ab] |
| **Monthly** | 38 | 0.03 (1.61) [bc] | 0.24 (0.85) [b] |
| **Rarely** | 190 | 0.88 (1.57) [c] | 0.48 (1.03) [c] |

[a.] Means not sharing any letters are significantly different by the Tukey test at the 5% level of significance.

## V. DISCUSSION

### A. Interpreting the Negotiation Indices and Their Distribution

The RBTI and CATI operationalize the core privacy negotiations theorized in youth engagement with SVAs. Their distributions provide an empirical map of how these tensions manifest across a population. The symmetric distribution of RBTI, centered near zero, reveals that youth are not universally risk or benefit dominant. A significant portion of youth reside in an intermediate zone where perceived risks and benefits are relatively balanced, embodying the everyday calculus of SVA engagement. For CATI, the mean near zero similarly indicates an overall balance between control and benefit-driven acceptance, but the slight positive skew reflects a small group of youth with particularly strong control-dominant profiles. At the same time, a modest majority of respondents occupy benefit-leaning positions (CATI < 0), underscoring that many youth resolve the control–acceptance tension in favor of convenience. Visualizing the joint distribution of RBTI and CATI (see Fig. 1) further clarifies the negotiation landscape. Respondents are relatively dispersed across all quadrants, confirming that youth view risk-benefit and control-acceptance evaluation in diverse ways. The largest portions exhibit either a benefit-focused acceptance (low RBTI, low CATI), where perceived benefits strongly outweigh both risks and sense of control, or demonstrate active concern (high RBTI, high CATI), where high perceived risks coincide with a strong relative sense of control and trust. These patterns validate that privacy attitudes are not rigid but are better represented as composite positions within a tension space, directly addressing RQ1 and RQ2.

### B. How Negotiating Tensions Predict Behavior

The relationship between the negotiation indices and PPB reframes the privacy paradox. Rather than a contradiction between concern and action, our analysis shows that behavior emerges from how underlying tensions are resolved. Both RBTI and CATI were positively correlated with PPB and remained significant predictors in a regression model controlling for age and SVA usage frequency. This suggests two pathways to protective behavior. First, a risk-dominant stance (high RBTI) promotes protection, consistent with privacy calculus models in which perceived threats motivate action [18]. Second, and more strongly in our model, a control-dominant stance (higher CATI) also promotes protection. Youth with higher privacy self-efficacy and trust are more likely to act, even when they also perceive benefits. This finding aligns with our prior structural model, which identified PSE as the strongest direct driver of PPB and showed that trust must operate through self-efficacy to enable action [15]. The privacy paradox can therefore be reinterpreted as a negotiation outcome. An individual who expresses high concern but takes no protective action may perceive high risks while lacking a sufficient sense of control, reflected in low self-efficacy and trust. In such cases, the tension is resolved through passive, benefit-driven acceptance rather than action. This "efficacy gap," where risk perception does not translate into self-efficacy, was quantified in our earlier work [15]. Thus, the paradox is clarified by modeling how competing tensions produce the observed behavioral pattern.

## C. Usage Frequency in Shaping Privacy Negotiations

SVA usage frequency emerged as a dynamic link in the privacy negotiation process. Our findings show that daily users have significantly lower RBTI and CATI scores than infrequent users, indicating that their negotiations are resolved into more benefit-driven, acceptance-leaning positions. Two interpretations follow. First, individuals who naturally weigh benefits more heavily or are more willing to accept defaults may be more inclined toward frequent use. This pattern aligns with the Technology Acceptance Model and diffusion of innovation theory, where high perceived usefulness and low privacy concern support adoption [27]. It is also consistent with privacy calculus, which suggests that individuals continuously weigh risks against benefits, making those with an initially benefit-skewed calculus more likely to become habitual users [18].

Second, repeated use and integration of SVAs into daily routines may gradually shift these indices over time, as users downplay risks, perceive greater benefits, and acclimatize to reduced control. Users may normalize the presence of SVAs, leading to greater trust and less desire for direct control, an acceptance supported by prior literature [3], [28]. This also mirrors automation trust research, which suggests that reliance develops over time and reduces tension between wanting control and surrendering it [29]. These explanations are not mutually exclusive, but likely form a reciprocal relationship in which initial convenience encourages frequent use, which then further reinforces benefit-driven acceptance and acclimatization.

Notably, frequent users reported lower CATI, meaning perceived benefits outweigh their sense of control. This suggests that high engagement does not necessarily increase self-efficacy or trust; instead, convenience may come at the cost of agency. This reinforces the design critique that SVA ecosystems often prioritize seamless function over transparent user control, fostering a form of resigned acceptance [15], [30].

## D. Implications for Privacy Design and Governance

The negotiation-based framing advanced in this paper has direct implications for SVA design and governance. The finding that control-dominant profiles (high CATI) are more strongly associated with privacy-protective behavior underscores the importance of supporting youth agency. Design interventions should therefore reduce friction in privacy management and make control visible and meaningful. Rather than treating privacy settings as secondary, SVA platforms could better support users' understanding of data flow and ability to intervene. This aligns with the negotiation perspective by recognizing that youth are not simply unconcerned about privacy but are balancing convenience against uncertainty and limited control. From a governance perspective, the indices developed in this study provide a tool for identifying populations particularly vulnerable to resigned acceptance. Youth in low-CATI positions may benefit from regulatory or educational interventions that emphasize not only risk awareness but also practical strategies for exercising control. Effective youth privacy protection therefore requires attention to both structural design choices and individual capability-building.

## E. Methodological Contribution of Negotiation Indices

This study contributes methodologically by demonstrating how negotiation indices can complement traditional construct-level models. RBTI and CATI do not replace existing measures of risk, benefit, trust, or self-efficacy. Instead, they place these constructs into interpretable dimensions that capture how competing considerations coexist within individuals. Such indices may be particularly useful in comparative research, longitudinal designs, or intervention studies, where shifts in negotiation profiles over time can be tracked. More broadly, this approach provides a practical way to capture ambivalence and negotiation in privacy decision-making while remaining grounded in established theories.

## F. Limitations & Future Work

While this study provides novel insights into the negotiation tensions of young SVA users, it has several limitations and clear directions for future research. First, the findings are based on a single sample using self-reported measures and a cross-sectional design. This limits our ability to make causal claims about the observed relationships or changes in attitudes over time. Second, the RBTI and CATI indices are linear combinations of our privacy constructs. While they are theoretically grounded and effective for initial exploration, alternative formulations could be tested to better capture the tension in these constructs. Finally, the reliance on self-reported data is inherently limited without behavioral or log data to corroborate reported usage patterns and privacy behaviors. Future studies should employ longitudinal or experimental designs to eliminate selection bias and validate RBTI and CATI indices across a diverse population. Furthermore, the indices themselves open new analytical research possibilities. They could be utilized with clustering techniques to create tension-based profiles of SVA users. Additionally, CATI and RBTI can be used as predictor variables in experiments, such as testing the efficacy of different privacy interfaces or transparency tools.

## VI. CONCLUSION

This study advances a negotiation-based understanding of youth privacy decision-making in smart voice assistant ecosystems. By introducing the risk-benefit tension index (RBTI) and the control-acceptance tension index (CATI), we demonstrate that youth privacy attitudes are not adequately captured by single constructs or linear predictors alone. Instead, privacy-protective behavior emerges from the resolution of competing pressures between perceived risks and benefits, and between a desire for control and the pull of convenience. Our findings show that both tensions matter, but that control-oriented negotiation plays an important role in enabling protective action. At the same time, frequent SVA use is associated with more benefit-dominant and acceptance-leaning profiles, suggesting that sustained engagement may normalize reduced control rather than strengthen it. These patterns help clarify the privacy paradox by revealing how high concerns can coexist with low action when negotiation resolves in favor of acceptance rather than agency. By reframing privacy decision-making as negotiation, this paper offers a conceptual and methodological contribution that complements existing models and provides a more nuanced lens for understanding youth experiences with SVAs. Future research can build on this approach by applying negotiation indices in other settings, across cultural contexts, or in experimental designs that test how changes in transparency and control reshape privacy negotiations over time. Ultimately, supporting youth privacy in SVA ecosystems requires recognizing and addressing the tensions that structure everyday digital decision-making, rather than assuming uniform concern or rational trade-off.

ACKNOWLEDGMENT

This project has been funded by the Office of the Privacy Commissioner of Canada (OPC); the views expressed are those of the authors and do not necessarily reflect those of the OPC.